\documentclass[12pt]{article}
\usepackage{amsmath,amsthm,amscd,amssymb}
\usepackage{latexsym}

\theoremstyle{plain}
\usepackage{epsf}
\usepackage{epsfig}
\newlength{\mycolspace}
\theoremstyle{definition}

\theoremstyle{remark}

\newcommand{\calH}{{\cal{H}}}

\newcommand{\calL}{{\cal{L}}}

\newcommand{\calR}{{\cal{R}}}

\newcommand{\calT}{{\cal{T}}}

\newcommand{\calX}{{\cal{X}}}

\newcommand{\fr}{{^F\hspace{-.02in}R}}
\newcommand{\fcalR}{{^F\hspace{-.02in}\calR}}
\newcommand{\fg}{{^F\hspace{-.04in}g}}
\newcommand{\fgam}{{^F\hspace{-.00in}\Gamma}}

\begin{document}
\title{The Generalized Jacobi Equation}
\author{C. Chicone\\Department of Mathematics\\University of
Missouri-Columbia\\Columbia, Missouri 65211, USA
\and B. Mashhoon\thanks{Corresponding author. E-mail:
mashhoonb@missouri.edu (B. Mashhoon).} \\Department of Physics and
Astronomy\\University of Missouri-Columbia\\Columbia, Missouri 65211, USA}
\maketitle
\begin{abstract}
The Jacobi equation in pseudo-Riemannian geometry determines the
linearized geodesic flow. The linearization ignores the relative
velocity of the geodesics.  The generalized Jacobi equation takes the
relative velocity into account; that is, when the geodesics are neighboring
but their relative velocity is arbitrary the corresponding geodesic
deviation equation is the generalized Jacobi equation. 
The Hamiltonian structure of this nonlinear equation is analyzed in this paper.
The tidal accelerations for test particles in the field of a plane
gravitational wave and the exterior field of a rotating mass are investigated.
In the latter case, the existence of an attractor of uniform relative radial motion
with speed $2^{-1/2}c\approx 0.7 c$ is pointed out. 
The astrophysical implication of this result for the terminal speed of a
relativistic jet is briefly explored. 
\end{abstract}
\section{Introduction}
The analysis of observations in Einstein's general relativity theory is
a rather complicated issue in contrast with the elegant simplicity
of the geometric structure of the theory~\cite{s3}. Take, for instance, the excess
perihelion precession of Mercury that was explained by Einstein
in 1915 and provided the first major success of general relativity.
This standard theoretical result follows from the solution
of the geodesic equation in the post-Newtonian approximation.
However, the relevant observational results are usually
obtained by monitoring the motion of Mercury from the Earth via
electromagnetic radiation that is reflected by Mercury and reaches astronomical
telescopes on the Earth. In effect, one has to deal with the \emph{relative}
motion of one geodesic (``Mercury'') with respect to the other
(``Earth'')~\cite{b1}. Ideally, the observer following the reference 
geodesic sets up in its neighborhood a laboratory, where the 
measuring devices are assumed to work as in inertial spacetime.
This local inertial frame of the observer is represented by the
Fermi coordinate system. To construct such a coordinate patch, an
orthonormal set of ideal gyroscopes would be required to define the local
spatial frame of the observer, while an ideal  standard clock would measure
the observer's proper time. The reduced equation of geodesic motion of 
a nearby particle in such a system is the generalized Jacobi equation
that will be studied in this paper. The results may be applied to 
high energy astrophysical phenomena as well as precision measurements
in space involving the tidal field of the Earth or an incident gravitational
wave.

In a congruence of timelike geodesics, neighboring curves have rates
of separation that are usually negligible compared to the speed of
light in vacuum $c$. In this case the geodesic deviation equation
is the standard Jacobi equation~\cite{bee}.
There are physical circumstances, however, where the worldlines of adjacent
geodesics diverge rapidly. Consider, for instance, a congruence of
test particles with a common initial velocity falling toward a 
black hole such that depending on their impact parameters, some
particles would be captured by the black hole,
while others would escape its gravitational pull. Though the geodesics are
neighboring as they approach the black hole, the rate of separation 
between the particles that fall in and those that are merely deflected
would no longer be small compared to $c$. This example makes it
clear that the relevant equation of relative motion, i.e. the
generalized Jacobi equation, has a limited domain of applicability,
since rapidly diverging geodesics would no longer be adjacent after
a period of time that is characteristic of the spacetime curvature---that is,
$\varrho/c$, where $\varrho$ is the radius of curvature.

The investigation of the relative motion of neighboring observers following
arbitrary timelike curves leads to the general deviation equation. 
The relativistic theory of tides is based on the general \emph{geodesic}
deviation equation. This is rather complicated when the rate of separation
of geodesics could be close to the speed of light; therefore, we confine
our detailed discussion here to a limited form of this equation that is valid
to first order in the relative separation (i.e.\ the generalized Jacobi equation).
On the other hand, in the (``nonrelativistic'') case where a smooth
geodesic congruence is given, a general treatment of the geodesic deviation
equation is available; in fact, 
a new derivation can be found in the recent
work of Kerner, van Holten and Colistete~\cite{k} who also
give a novel application: they use the geodesic deviation equation
together with Poincar\'e's perturbation method to 
approximate bounded (eccentric) geodesics in the Schwarzschild metric that are
close to circular geodesics. 

The generalized Jacobi equation is discussed in detail in Section~2;
in particular, we are interested in the dynamical system represented
by this nonlinear equation, where the nonlinearity is due to the
existence of velocity-dependent terms in the system. 
Therefore, we present a detailed analysis of its
Hamiltonian structure.

To illustrate certain general dynamical features of the generalized
Jacobi equation, we consider the relative motion of neighboring test
particles in the field of a plane gravitational wave in Section~3
and in the exterior field of a gravitating source in Section~4.
We note that many compact astrophysical systems emit powerful 
oppositely-directed jets that are highly collimated. 
No basic theory of astrophysical jets is available
at present, though many---especially magnetohydrodynamic---models have
been proposed. In particular, the physical processes that determine the 
asymptotic speed of the bulk flow are not clear at this time.
Of particular interest in this connection 
is the relevance of the relative radial motion
described in Section~4 to the ``terminal'' bulk motion of a relativistic
jet once such an outflow is beyond the active region around the source
and the bulk motion is subject only to the gravitational attraction of
the source. Taking only the relativistic tidal forces into account,
we demonstrate in this case  the presence of an ``attractor''
at a Lorentz factor $\Gamma=\sqrt{2}$, corresponding to 
$V=c/\sqrt{2}\approx 0.7c$, for the terminal bulk motion of a jet relative to
the ambient medium.

Finally, Section~5 contains a discussion of our results. In the following
sections, we use units such that $c=1$. 
 
\section{Generalized Jacobi equation}
The generalized Jacobi equation in Fermi coordinates can be simply
obtained by writing an alternative form of the geodesic equation, i.e.
the \emph{reduced} geodesic
equation~\eqref{geoiso} below, in Fermi coordinates. However, to 
clarify the Hamiltonian and Lagrangian aspects of this equation, it is
useful to proceed as in the following subsections.
\subsection{Hamiltonian and Lagrangian formalisms}
As is well-known, the geodesic equation in local
coordinates $x^\mu$ on a pseudo-Riemannian manifold with metric 
\[ds^2=g_{\mu\nu}(x) dx^\mu dx^\nu\]
is equivalent to the
Euler-Lagrange equation for the Lagrangian  
\[\calL (x^\mu, u^\nu)=\frac{1}{2}\, g_{\mu\nu}(x) u^\mu u^\nu,\]
where 
\[u^\mu:=\frac{dx^\mu}{d\lambda},\]
the canonical momentum is given by 
\[p_\mu=\frac{\partial \calL}{\partial u^\mu}=g_{\mu\nu}(x)u^\nu,\]
and $\lambda$ is an arbitrary parameter (see~\cite{am}). In fact,
the Euler-Lagrange equation for this system is equivalent to
\[
\frac{d^2x^\mu}{d\lambda^2}
+\Gamma^\mu_{\alpha\beta} 
\frac{dx^\alpha}{d\lambda} \frac{dx^\beta}{d\lambda}=0,
\]
where 
\[
\Gamma^\mu_{\alpha\beta}=
\frac{1}{2} g^{\mu\gamma}(g_{\gamma\alpha,\beta}+g_{\gamma\beta,\alpha}
-g_{\alpha\beta,\gamma})
\]
are the Christoffel symbols.
The corresponding Hamiltonian is
\[
\calH(x^\mu,p_\nu)=p_\mu u^\mu-\calL(x^\mu, u^\nu)
=\frac{1}{2} g^{\mu\nu}(x) p_\mu p_\nu, 
\]
which is independent of the parameter $\lambda$;
thus, $\calH$ is constant on geodesics. It follows that along
a geodesic $\lambda$ is a linear function of $s$. 

\subsection{Isoenergetic reduction}
Let us specialize to a four-dimensional Lorentzian manifold where we take 
an admissible coordinate chart and the
sign convention $(-,+,+,+)$ for the Lorentzian metric. 
We will also use
the Einstein summation rules; Greek indices run from zero to three and Latin
indices run from one to three. Moreover, round brackets around indices
denote symmetrization, while square brackets denote antisymmetrization. 

The timelike geodesics ($(ds/d\lambda)^2<0$) with energy $\calH=-1/2$ have
$\lambda=\tau$, where $\tau$ is the 
proper time along the geodesic such that $(ds/d\tau)^2=-1$.
We will determine the isoenergetic reduction for the Hamiltonian system
restricted to this manifold. 

The Hamiltonian equations of motion (where we have assumed for 
simplicity that $p^\mu=u^\mu$; that is, the mass of the free test
particle is normalized to unity) 
are given by
\begin{eqnarray}\label{hamsys}
\nonumber \frac{dx^\mu}{d\tau}&=&\frac{\partial \calH}{\partial p_\mu}
   =g^{\mu\nu} p_\nu=p^\mu,\\
\frac{dp_\mu}{d\tau}&=&-\frac{\partial \calH}{\partial x^\mu}
=-\frac{1}{2} g^{\alpha\beta}_{\hspace{.14in},\mu}\, p_\alpha p_\beta .
\end{eqnarray}  
Reduction is accomplished by first separating the pairs of 
canonically conjugate
variables $(x^\mu, p_\nu)$ into time and space variables; that is, we 
take 
$(x^\mu, p_\nu)=(t,p_0; x^i,p_j)$ so that  points in spacetime with 
energy $-1/2$ 
satisfy the equation
\begin{equation}\label{eneq}
\calH(t,p_0; x^i,p_j)=\frac{1}{2} g^{00}p_0^2+g^{0i}p_0p_i
+\frac{1}{2} g^{ij}p_ip_j=-\frac{1}{2}.
\end{equation}
Because the flow of time does not stop along the worldline of an observer,
we have $dt/d\tau\ne 0$. Hence, by Hamilton's equations, the partial 
derivative  
${\partial\calH}/{\partial p_0}$ does not vanish. By the implicit
function theorem, there is a function $\alpha$ such that
\begin{equation}\label{hhalf}
\calH(t,\alpha(t,x^i,p_j),x^i,p_j)=-\frac{1}{2}.
\end{equation} 
That is,
the energy surface is (locally) the graph of $\alpha$.
Of course, we can also solve explicitly for $\alpha$. Indeed, because 
$g^{00}\ne 0$ in an admissible chart we can complete
the square with respect to $p_0$ in
equation~\eqref{eneq} to obtain
\[
\big(p_0+\frac{g^{0i}p_i}{g^{00}}\big)^2=
-\frac{1+\tilde g^{ij}p_ip_j}{g^{00}}\,,
\]
where 
\[ 
\tilde g^{ij}:=g^{ij}-\frac{g^{0i}g^{0j}}{g^{00}}\,.
\]
Hence, 
\[
\alpha(t,x^i,p_j)=-\frac{g^{0i}}{g^{00}}p_i
\pm \Big(\frac{1+\tilde g^{ij}p_ip_j}{(-g^{00})}\Big)^{1/2}.
\]
We note that $(\tilde g^{ij})$ is the inverse of the 
spatial metric $(g_{ij})$; that is, $\tilde g^{ik}g_{kj}=\delta^i_j$.  

On the energy surface,  the equation
\[
\frac{dp_0}{d\tau}=
-\frac{\partial \calH}{\partial t}(t,\alpha(t,x^i,p_j),x^i,p_j)
\]
decouples from the system~\eqref{hamsys}. 
By computing the partial derivatives with respect to $x^i$ and $p_i$
on both sides of equation~\eqref{hhalf}, we obtain the identities
\begin{eqnarray*}
\frac{\partial\calH}{\partial p_0} \frac{\partial\alpha}{\partial x^i}
+ \frac{\partial\calH}{\partial x^i} &=& 0,\\
\frac{\partial\calH}{\partial p_0} \frac{\partial\alpha}{\partial p_i}
+ \frac{\partial\calH}{\partial p_i} &=& 0.
\end{eqnarray*}
Using these relations, the Hamiltonian system~\eqref{hamsys} for the
spatial variables can be expressed as
\begin{eqnarray*}
\frac{dx^i}{d\tau}&=&\frac{\partial\calH}{\partial p_i}
=-\frac{\partial\calH}{\partial p_0}\frac{\partial\alpha}{\partial p_i},\\
\frac{dp_i}{d\tau}&=&-\frac{\partial\calH}{\partial x^i}
=\frac{\partial\calH}{\partial p_0}\frac{\partial\alpha}{\partial x^i},
\end{eqnarray*}
where $\partial\calH/\partial p_0=d t/d\tau\ne 0$ since
$t$ is the timelike variable.
Hence, the remaining ``spatial'' equations in the Hamiltonian 
system~\eqref{hamsys} can
be recast into the time-dependent Hamiltonian system (``the isoenergetic
reduction'')
\begin{eqnarray}\label{hamiso}
\nonumber\frac{dx^i}{dt}&=&-\frac{\partial \alpha}{\partial p_i}(t,x^j,p_k),\\
\frac{dp_i}{dt}&=&\frac{\partial \alpha}{\partial x^i}(t,x^j,p_k),
\end{eqnarray} 
with Hamiltonian $-\alpha$.

The Hamiltonian isoenergetic reduction procedure  is equivalent to
a Lagrangian procedure. In fact, let us define the Lagrangian~\cite{b1}
\[
L(t,x^i,v^j)=
-\big[-(g_{00}(t,x)+2 g_{0i}(t,x) v^i+g_{ij}(t,x) v^iv^j)\big]^{1/2}, 
\]
where $v^i:=dx^i/dt$, the minus sign under the square root 
ensures that $L$ is real
and the other minus sign is to 
conform with the sign convention used in the 
Hamiltonian system~\eqref{hamiso}.
We now have 
\[ p_i:=\frac{\partial L}{\partial v^i}=
(g_{0i}+g_{ij}v^j)
\big[-(g_{00}+2 g_{0i} v^i+g_{ij} v^iv^j)\big]^{-1/2}
\]
and the corresponding time-dependent Hamiltonian 
\[H=H(t,x^i,p_j):=p_i v^i-L(t,x^i,v^j)\]
is given by
\begin{eqnarray*}
H &=& -(g_{00}+g_{0i}\frac{ dx^i}{dt}) \frac{dt}{d\tau}\\
&=& -(g_{00}p^0+g_{0i}p^i)=-g_{0\alpha}p^\alpha=-p_0,
\end{eqnarray*}
where we have used the fact that the proper time $\tau$ along the
geodesic worldline is such that $L=-d\tau/dt$.
Hence $H=-\alpha$ in system~\eqref{hamiso}.

This reduced system is equivalent to the second-order differential equation
\begin{equation}\label{geoiso}
\frac{d^2x^i}{dt^2}
-(\Gamma^0_{\alpha\beta}\frac{dx^\alpha}{dt}\frac{dx^\beta}{dt})\frac{dx^i}{dt}
+\Gamma^i_{\alpha\beta}\frac{dx^\alpha}{dt}\frac{dx^\beta}{dt}=0,
\end{equation}
which is the reduced geodesic equation.
To prove this fact, note that the Hamiltonian or Lagrangian equations
of motion are equivalent to  the Euler-Lagrange equation
\begin{eqnarray}\label{eleq}
\lefteqn
\nonumber\lefteqn{\frac{d}{dt}\big[(g_{0i}+g_{ij}v^j)
[-(g_{00}+2 g_{0i} v^i+g_{ij} v^iv^j)]^{-1/2} \big]}\\
&& =\frac{1}{2}(g_{00,i}+2 g_{0j,i} v^j+g_{kl,i} v^kv^\ell)
[-(g_{00}+2 g_{0i} v^i+g_{ij} v^iv^j)]^{-1/2}.
\end{eqnarray}
Let us use the proper time $\tau$ to recast equation~\eqref{eleq} 
in the form
\[
\frac{d}{d\tau}(g_{i\alpha}u^\alpha)
=\frac{1}{2}g_{\alpha\beta,i}u^\alpha u^\beta,
\]
where $u^{\mu}=dx^{\mu}/d\tau$.
After computation of the derivative by the product rule, the equation
can be rearranged to the equivalent form
\[
g_{i\alpha}\frac{du^\alpha}{d\tau}+\frac{1}{2} 
(g_{i\alpha,\beta}+g_{i\beta,\alpha}-g_{\alpha\beta,i})u^\alpha u^\beta=0.
\]
We multiply both sides of the last equation by $g^{\mu i}$ and use the
identity
$g^{\mu i}g_{i\alpha}+g^{\mu 0}g_{0\alpha}=\delta^\mu_\alpha$
to show that
\[
(\delta^\mu_\alpha-g^{\mu 0}g_{0\alpha})\frac{du^\alpha}{d\tau}
+\frac{1}{2}g^{\mu i} 
(g_{i\alpha,\beta}+g_{i\beta,\alpha}-g_{\alpha\beta,i})u^\alpha u^\beta=0.
\]
Also, using the identity
\[
\Gamma^\mu_{\alpha\beta}
=\frac{1}{2}g^{\mu i} 
(g_{i\alpha,\beta}+g_{i\beta,\alpha}-g_{\alpha\beta,i})
+\frac{1}{2}g^{\mu 0} 
(g_{0\alpha,\beta}+g_{0\beta,\alpha}-g_{\alpha\beta,0}),
\]
we have the equation
\begin{equation}\label{eq:el2}
\frac{du^\mu}{d\tau} + \Gamma^\mu_{\alpha\beta} u^\alpha u^\beta
-g^{\mu 0}g_{0\alpha} \frac{du^\alpha}{d\tau}
 -\frac{1}{2} g^{\mu 0} 
(g_{0\alpha,\beta}+g_{0\beta,\alpha}-g_{\alpha\beta,0})u^\alpha u^\beta=0.
\end{equation}

We claim that 
\begin{equation}\label{eq:ne8}
u_\mu(\frac{du^\mu}{d\tau} + \Gamma^\mu_{\alpha\beta} u^\alpha u^\beta)=0.
\end{equation}
To prove this identity, we replace 
$\Gamma^\mu_{\alpha\beta}$ by its defined value, 
multiply through by $u_\mu$ and then
use the antisymmetry of the last two terms in the resulting expression
to conclude that
\[ 
u_\mu(\frac{du^\mu}{d\tau} + \Gamma^\mu_{\alpha\beta} u^\alpha u^\beta)
=u_\mu\frac{du^\mu}{d\tau}
+\frac{1}{2} u^\rho g_{\rho\alpha,\beta} u^\alpha u^\beta.
\]
A calculation shows that the right-hand side of this last equation
is equal to 
\[
\frac{1}{2}\frac{d}{d\tau} (g_{\mu\nu} u^\mu u^\nu),
\] 
and therefore it is equal to zero, since 
$(ds/d\tau)^2=g_{\mu\nu}u^\mu u^\nu=-1$.
 
After multiplying equation~\eqref{eq:el2} by $u_\mu$, 
we use equation~\eqref{eq:ne8} to see that
\[
g_{0\alpha}\frac{du^\alpha}{d\tau}+\frac{1}{2}
(g_{0\alpha,\beta}+g_{0\beta,\alpha}-g_{\alpha\beta,0})u^\alpha u^\beta=0,
\]
since $u_\mu g^{\mu 0}=u^0=dt/d\tau\ne 0$.
By inserting this equality into equation~\eqref{eq:el2},
it follows that 
\[
\frac{du^\mu}{d\tau}+\Gamma^\mu_{\alpha\beta}u^\alpha u^\beta=0.
\]
In particular, if $\mu=0$, then
\begin{equation}\label{teq}
\frac{d^2 t}{d\tau^2}+\Gamma^0_{\alpha\beta}u^\alpha u^\beta=0;
\end{equation}
and if $\mu=i$, then
\begin{equation}\label{xeq}
\frac{d^2x^i}{d\tau^2}+\Gamma^i_{\alpha\beta}u^\alpha u^\beta=0.
\end{equation}

Using the chain rule, we have that
\[
\frac{d^2x^i}{d\tau^2}=\big(\frac{dt}{d\tau}\big)^2\,\frac{d^2x^i}{dt^2}
+\frac{d^2t}{d\tau^2}\frac{dx^i}{dt}.
\]
By inserting this identity into equation~\eqref{xeq} and by replacing
$dx^\rho/d\tau$ with $(dx^\rho/dt)(dt/d\tau)$, we obtain 
equation~\eqref{geoiso} with its left-hand side 
multiplied by the nonzero factor
$\big({dt}/{d\tau}\big)^2$.
This proves the desired result.
\subsection{Generalized Jacobi equation}
The generalized Jacobi equation is the linearization of the reduced
geodesic equation~\eqref{geoiso} in Fermi coordinates
with respect to the space variables, but not their velocities. 
We will obtain the explicit expression for this equation in 
Fermi coordinates valid in a cylindrical
spacetime region along the worldline of a reference observer, 
which is at rest at 
the spatial origin of the Fermi coordinates (see Appendix~A). 
This is perhaps the most useful form of the
generalized Jacobi equation. 

For an arbitrary spacetime,
when a reference observer (that follows
a geodesic) in a coordinate chart $(t,x^i)$ is specified, 
we can always find a Fermi coordinate system
$(T,X^i)$ in a neighborhood of 
the worldline of the fiducial observer so that the metric tensor is given by
\begin{eqnarray}
\label{metrict} 
\nonumber g_{00}&=&-1-\fr_{0i0j}(T)X^iX^j+\cdots,\\
\nonumber g_{0i}&=&-\frac{2}{3}\,\fr_{0jik}(T) X^jX^k+\cdots,\\
g_{ij}&=&\delta_{ij}-\frac{1}{3}\,\fr_{ikj\ell}(T) X^kX^\ell+\cdots,
\end{eqnarray}
where 
\[
\fr_{\alpha\beta\gamma\delta}
=R_{\mu\nu\rho\sigma}\lambda^\mu_{(\alpha)}\lambda^\nu_{(\beta)}
\lambda^\rho_{(\gamma)}\lambda^\sigma_{(\delta)},
\]
are the  components of the Riemann tensor projected onto the 
orthonormal tetrad frame $\lambda^\mu_{(\alpha)}(T)$ that is
parallel propagated
along the worldline of the reference observer at rest at the spatial 
origin of the
Fermi chart (cf.\ Appendix~A). By expanding the coefficients of the
reduced geodesic equation~\eqref{geoiso} for the metric expressed
in the Fermi coordinates and retaining only the first-order terms
in the spatial directions, we obtain the generalized Jacobi equation
\begin{eqnarray}\label{gje}
\nonumber 
\lefteqn{\frac{d^2X^i}{dT^2}+\fr_{0i0j}X^j +2\,\fr_{ikj0}V^k X^j}\\
&&+(2\,\fr_{0kj0} V^iV^k+\frac{2}{3}\,\fr_{ikj\ell}V^kV^\ell
+\frac{2}{3}\,\fr_{0kj\ell}V^iV^kV^\ell) X^j=0,
\end{eqnarray}
where $V^i:=dX^i/dT$. Alternatively, we could start from the Hamiltonian
equations~\eqref{hamiso} and substitute in the Hamiltonian for
$g^{00}$, $g^{0i}$ and $\tilde g^{ij}$ expansions corresponding to 
equations~\eqref{metrict} and recover equation~\eqref{gje} directly
from Hamilton's equations~\eqref{hamiso}; this is done in Appendix~B.
The generalized Jacobi equation in arbitrary spacetime coordinates is given
in Appendix~C.

It is interesting to note a special scaling property of equation~\eqref{gje}.
To this end, we let $\varrho$ denote the radius of curvature; it is defined here
so that $1/\varrho^2$ is the supremum of the absolute
magnitudes of the components
of the Riemann tensor over the coordinate patch where the Fermi coordinates
are defined. We will use the  natural rescaling of the generalized 
Jacobi equation given by 
\begin{equation}\label{natre}
X^i=\varrho \hat x^i,\qquad T=\varrho \hat t.
\end{equation}
In fact, under this rescaling the generalized Jacobi equation maintains
its form except that each component of the Riemann tensor
is multiplied by $\varrho^2$. The scaled equation is valid
for $|\hat x^i|\ll 1$ and for arbitrary velocities  whose magnitudes do 
not exceed
the speed of light $c=1$.  The requirement that $|\hat{\mathbf{x}}|\ll 1$
would naturally imply that in most cases the generalized Jacobi equation
would hold for a limited interval of time $\hat t-\hat t_0$
 starting from some initial state
at $\hat t=\hat t_0$ such that $|\hat t-\hat t_0|\lesssim 1$.

The Hamiltonian dynamics of the generalized Jacobi equation will be 
explored in the following section for relative motion in the field
of a plane gravitational wave. The dimensionless Fermi coordinates
$(\hat t,\hat {\mathbf{x}})$ introduced in equation~\eqref{natre} 
will only be employed
in the latter part of Section~3 and should not be confused with the usual
local coordinates that are used to express the spacetime metric of the plane wave 
below.

\section{Generalized Jacobi equation for a plane gravitational wave}
In this section we will determine the generalized Jacobi equation for
a \emph{plane} gravitational wave given by the metric 
\[
ds^2=-dt^2+dx^2+U^2(e^{2h}dy^2+e^{-2h}dz^2),
\]
where $U$ and $h$ are functions of retarded time $u=t-x$. This metric is 
a solution of Einstein's field equations in vacuum 
($R_{\mu\nu}=0$) provided that 
\begin{equation}\label{comcon}
U_{,uu}+(h_{,u})^2 U=0. 
\end{equation}
Under this assumption, the gravitational field is of Petrov type $N$ 
(see~\cite{bpr}) and represents a plane wave propagating along the
positive $x$-axis.  

When $|h|\ll 1$ and $U=1$,  the metric reduces to a linear plane
gravitational wave in the $T$-$T$ gauge that is linearly polarized
(``$\oplus$'').

Any observer that remains at rest at spatial coordinates $(x,y,z)$
follows a geodesic of this spacetime. This is a consequence of a more
general result discussed in Appendix~D. 

Let us consider the reference observer at rest at the spatial origin.
We claim that the orthonormal  tetrad  frame of this observer
\begin{eqnarray}\label{tframe}
\nonumber \lambda^\mu_{(0)}= (1,0,0,0),&\qquad & \lambda^\mu_{(1)}= (0,1,0,0),\\
\lambda^\mu_{(2)}= (0,0,U^{-1}e^{-h},0),&\qquad & 
\lambda^\mu_{(3)}= (0,0,0,U^{-1}e^{h})
\end{eqnarray}
is parallel propagated along its worldline; that is,
\[
\frac{d\lambda^\mu_{(\alpha)}}{dt} +\Gamma^\mu_{\rho\sigma}
 \lambda^\rho_{(0)}\lambda^\sigma_{(\alpha)}=0.
\]
It follows from Appendix~D that the worldline of this 
observer lies on a geodesic
(that is, the last equation holds for $\alpha=0$),
so it suffices to show that the  equation holds for $\alpha=i$.
In fact, because
the nonzero components of $\Gamma^i_{0\alpha}$ are
\[
\Gamma^2_{02}=\frac{1}{2}\frac{g_{22,0}}{g_{22}},\quad
\Gamma^3_{03}=\frac{1}{2}\frac{g_{33,0}}{g_{33}},
\]
it is easy to see that $\lambda^\mu_{(1)}$ satisfies the equation. The desired
result for 
$\lambda^\mu_{(2)}$ and $\lambda^\mu_{(3)}$ follows from the fact
that $U^{-1} \exp (-h)$ and $U^{-1} \exp (h)$ in equation~\eqref{tframe}
are respectively equal to 
$(g_{22})^{-1/2}$ and $(g_{33})^{-1/2}$ up to a sign factor.

To obtain the generalized Jacobi equation for the reference observer,
we must compute the Fermi components of curvature along its
worldline. Since the observer is at rest at the origin ($x=y=z=0$), 
these components will
be functions of its proper time $t=T$ only. In fact, for
\[
K(u):=h_{,uu}+2 h_{,u}\frac{U_{,u}}{U}
\]
we have that 
\begin{eqnarray*}
\fr_{1212}=\fr_{1220}=\fr_{2020}&=&-K(T),\\
\fr_{1313}=\fr_{1330}=\fr_{3030}&=& K(T)
\end{eqnarray*} 
are the only nonzero components except for the symmetries of the
Riemann tensor. In the Fermi system, the spatial coordinates $\mathbf{X}$
are characterized invariantly with respect to the tetrad system as 
described in Appendix~A.

By inserting the curvature components into the generalized Jacobi
equation~\eqref{gje} with $\dot X^i=dX^i/dT$, we have the following
system of second-order differential equations:
\begin{eqnarray}\label{thesys}
\nonumber \ddot X&=&-\frac{2}{3} K(T)(\dot X-1)
[X(\dot Y^2-\dot Z^2)
   -Y \dot Y(\dot X-3)\\
       \nonumber &&{}+Z\dot Z(\dot X -3)],  \\
\nonumber \ddot Y&=&-\frac{1}{3} K(T)\{
   2 X\dot Y ( \dot X+\dot Y^2-\dot Z^2)
     -Y[3+2 \dot Y ^2 (\dot X-3)-6 \dot X+2 \dot X ^2]\\
            \nonumber &&{}+2 Z\dot Y\dot Z(\dot X -3)\}, \\
\nonumber \ddot Z&=&\frac{1}{3} K(T)\{
    2 X\dot Z (\dot X-\dot Y ^2+\dot Z ^2)+2 Y\dot Y\dot Z(\dot X -3)\\
      &&{}-Z[3+2\dot Z ^2(\dot X-3)-6\dot X+2 \dot X ^2]\}.
\end{eqnarray}

We note that this system is invariant under the transformation
$X\mapsto X$, $Y\mapsto Z$, $Z\mapsto Y$ and $K\mapsto -K$.
The generalized Jacobi system~\eqref{thesys} with its velocity-dependent
nonlinearities may conceivably have practical applications in the future
for the case of space-based gravitational wave detectors 
such as LISA~\cite{lisa}.
\subsection{Dynamics}
System~\eqref{thesys} has a seven-dimensional extended state space such that
the five-dimensional submanifolds
\begin{eqnarray*}
M_Y&:=&\{(X,Y,Z,\dot X,\dot Y,\dot Z, T): Y=\dot Y=0\},\\
M_Z&:=&\{(X,Y,Z,\dot X,\dot Y,\dot Z,T): Z=\dot Z=0\}
\end{eqnarray*} 
are invariant. The extended state space is simply the space of
positions, velocities and time. 
On the three-dimensional invariant intersection
of these manifolds, the flow is given by 
$\ddot X=0$ and $\dot T=1$.
That is, equations~\eqref{thesys} permit the uniform motion of
a particle along the direction of wave propagation relative to the fiducial
particle. This is due to the transverse character of the gravitational
wave.
Since the corresponding two-parameter family of solutions is 
$X(T)=\dot X(0) T+X(0)$,
the relative motion is always unstable. That is,
most  members of this two-parameter family of geodesics diverge from the 
reference geodesic, which is the member of this family with 
$X(0)=\dot X(0)=0$.

In the small-$h$ approximation for a monochromatic wave with $U=1$, 
the curvature is given by $K=h_{,uu}$; hence, the assumption that 
$h(u)=-\epsilon \cos(\phi+\Omega u)$
results in the 
curvature $K(T)=\epsilon \Omega^2 \cos(\phi+\Omega T)$, where 
$\epsilon\ll 1$, $\Omega$ is the wave frequency and $\phi$ is a constant
phase. In this case, the radius of curvature is
$\varrho=1/(\Omega \sqrt{\epsilon})$ and, 
using the scaled variables~\eqref{natre},
system~\eqref{thesys} maintains its form except that
$K(T)$ is replaced by $k(\hat t):=\cos(\phi+\hat t/\mu)$, where
$\mu:=\sqrt{\epsilon}$.
It is convenient therefore to define a new temporal variable
$S:=\hat t/\mu$ and consider the equivalent first-order system with   
\[
\hat x'=\mu \xi,\quad \hat y'=\mu \eta, \quad \hat z'=\mu \zeta,
\]
where  a prime denotes differentiation with respect to the
new ``fast'' time $S$.
Each component of the vector field corresponding to the 
resulting first-order system has
order $\mu$ and is $2\pi$-periodic. 
This system is therefore in the correct form for averaging with respect to 
the temporal variable.

By averaging to second-order in the small parameter $\mu$, we transform
system~\eqref{thesys} (by a near-identity nonautonomous change of 
variables) to a system that is autonomous to second-order in $\mu$. 
In fact, the second-order approximation of the averaged system is given by
\begin{eqnarray}\label{avesys}
\nonumber \hat x'&=&\mu \xi,\\
\nonumber \xi'&=&0,\\
\nonumber \hat y'&=&\mu \eta-\mu^2 y\sin\phi,\\
\nonumber \eta'&=&\mu^2\eta\sin\phi,\\
\nonumber \hat z'&=&\mu \zeta+\mu^2 z\sin\phi,\\
          \zeta'&=&-\mu^2\zeta\sin\phi.
\end{eqnarray}
This second-order approximation is not affected
by the nonlinear terms in the generalized Jacobi equation; that is, the result
would be the same for the corresponding Jacobi equation. The effect of
the nonlinearities first appears in the third-order averaged system, i.e.
at order $\epsilon^{3/2}$. 

Let us note that taking into account the 
first-order approximation in $\mu$ only, 
the approximation procedure suggests
that on average $X$, $Y$ and $Z$ are not affected by the presence of
the wave; that is, to first order in $\mu$,
$\mathbf{X}(T)=\mathbf{X}(0)+\dot{\mathbf{X}}(0) T$. 
In case $\sin\phi\ne 0$, the first correction to this
approximation---necessary only for $Y(T)$ and $Z(T)$---is obtained 
from the second-order approximation of the averaged
system~\eqref{avesys}; in fact, we
have that
\begin{eqnarray}\label{eq:avans}
\nonumber Y(T)&=&Y(0) e^{-\epsilon\Omega T\sin\phi} 
  +\frac{\dot Y(0)}{\epsilon \Omega \sin\phi}\sinh(\epsilon \Omega T\sin\phi)\\
\nonumber &=&Y(0)+\dot Y(0) T-\epsilon Y(0)\Omega T\sin\phi+O(\epsilon^2),\\
\nonumber Z(T)&=&Z(0) e^{\epsilon\Omega T\sin\phi} 
  +\frac{\dot Z(0)}{\epsilon \Omega \sin\phi}\sinh(\epsilon \Omega T\sin\phi)\\
&=&Z(0)+\dot Z(0) T+\epsilon Z(0)\Omega T\sin\phi+O(\epsilon^2).
\end{eqnarray}
It follows from these results that for
$\sin\phi= 0$, the motion is on average unaffected by the presence
of the wave to first order in $\epsilon$, as expected from the form
of equation~\eqref{avesys}. Moreover, equations~\eqref{eq:avans}
ignore nonlinearities in the system~\eqref{thesys} and 
are therefore valid for the Jacobi equation averaged to order $\epsilon$.
It should be noted that for the 
Jacobi equation the motions in $Y$ and $Z$ are stable as follows from an
analysis of the corresponding 
Mathieu equations. This is not in conflict with equations~\eqref{eq:avans}
since to see this stability one has to average to higher orders than
$\mu^2=\epsilon$.

\subsection{Hamiltonian Chaos}\label{sec:hc}
The dynamical behavior of the generalized Jacobi system~\eqref{thesys} is
complicated by its high dimensionality and the velocity-dependent
nonlinear terms. On the other
hand, the system is Hamiltonian and if $K$ is periodic, 
we  expect some dynamical effects associated with Hamiltonian chaos:
resonance phenomena, stochastic regions and diffusion. While an investigation
of these phenomena is beyond the scope of this paper, we illustrate
the existence of Hamiltonian chaos in generalized Jacobi equations for
the simpler system with $1+1/2$ degrees of freedom given by 
\begin{equation}\label{eq:2djac}
\frac{d^2 X}{dT^2}=K(T)(1-2 V^2) X,
\end{equation}
where $V:=dX/dT$ and $K$ is a periodic function. This equation of 
relative motion
is obtained from the generalized Jacobi equation by limiting
attention to a two-dimensional world, i.e.\ a surface $(T,X)$, where $K$ is
the Gaussian curvature of the surface (cf. Section~4).
According to the results of Appendix~B, equation~\eqref{eq:2djac}
can be obtained from a Hamiltonian
\[ H(T,X,p)=\mp (1+p^2)^{1/2}[1+\frac{1}{2} K(T) X^2],\]
once only terms linear in $X$ are retained in the equation of relative motion.

It is interesting to remark that the equivalent first-order system 
\begin{equation}\label{eq:2djacfo}
\dot X=V,\qquad 
 \dot V=K(T)(1-2 V^2) X,
\end{equation}
\begin{figure}[ht]
\centerline{\epsffile{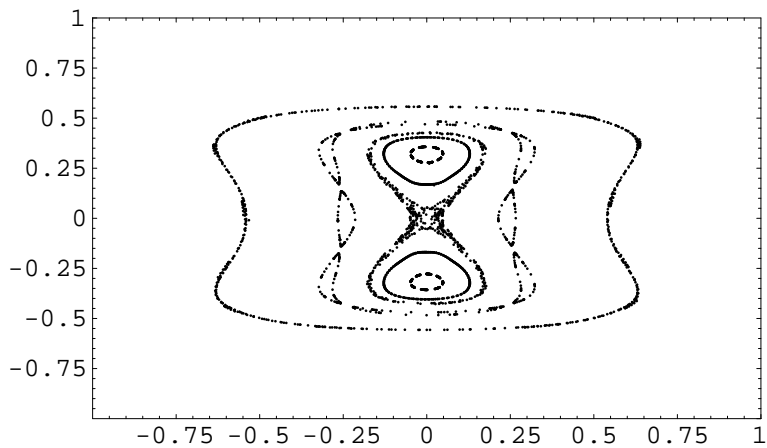}}
\caption{The plot of $V$ versus $X$ at $T=2n\pi$ for $n=0,1,2,\ldots,600$
is depicted for several orbits of the
Poincar\'e map for the system~\eqref{eq:2djac} with $K(T)=0.5 \cos (T)$.
\label{fig0}}
\end{figure} 
is Hamiltonian  with respect to the symplectic form
\[\omega=(1-2V^2)^{-1/2}dX\wedge dV,\]
in the region of the state space where $V^2\ne 1/2$. Also, 
we note that the change of coordinates 
\[
x=X,\qquad v=\frac{1}{2\sqrt{2}}\ln\Big(\frac{1+\sqrt{2} V}{1-\sqrt{2}V}\Big),
\]
for $V^2\ne 1/2$, 
transforms system~\eqref{eq:2djacfo} to 
\[
\dot x=\frac{1}{\sqrt{2}}
\Big(\frac{\exp(2\sqrt{2}\,v)-1}{\exp(2\sqrt{2}\,v)+1}\Big),
\qquad
\dot v=-K(T) x,
\]
a system that is Hamiltonian with respect to the usual symplectic structure.

A few typical orbits with $|V|<(1/2)^{1/2}$ are illustrated
in Figure~\ref{fig0} for a stroboscopic Poincar\'e map 
obtained
by numerical integration of system~\eqref{eq:2djacfo} with $K(T)=0.5 \cos (T)$; 
that is, we plot $V$ versus $X$ at times
$2n\pi$ for nonnegative integer values of $n$. 
We note the presence of stochastic layers,
resonant island chains and invariant closed curves. 
The proof of the existence of chaotic invariant sets even in this
two-dimensional case is not simple since a straightforward 
application of the Melnikov method runs into standard difficulties;
in particular, one has to deal with the problem of exponentially small
splittings of homoclinic loops.

\section{Generalized Jacobi equation for radial motion in black hole
spacetimes}
Imagine the relative motion of nearby free test particles along
the axis of rotational symmetry of a Kerr source.
For the reference geodesic, the equation of motion, using the
standard Boyer-Lindquist radial coordinate $r$, is~\cite{b1}
\begin{equation}\label{eq:4.1}
\big(\frac{dr}{d\tau}\big)^2=\gamma^2-1+\frac{2GMr}{r^2+a^2},
\end{equation}
where $G$ is Newton's gravitational constant, 
$M$ is the mass of the Kerr source and $aM$ is its
angular momentum.  Here $\gamma$ is a positive constant of
integration and $\gamma>1$ has the interpretation of the Lorentz factor
for the particle at infinity ($r\to\infty$).

For a neighboring geodesic, the generalized Jacobi equation
describing motion along the axis relative to the fiducial observer
is 
\begin{equation}\label{eq:4.2}
\frac{d^2R}{dT^2}+k(T)\big[1-2\big(\frac{dR}{dT}\big)^2\big]R=0,
\end{equation}
where $\mathbf{X}=(R,0,0)$, $R$ is the relative radial distance along the
axis, $T=\tau$ along the reference geodesic
and $k(T)$ is obtained~\cite{b1} from the solution of
equation~\eqref{eq:4.1} and
\begin{equation}\label{eq:4.3}
k=-2\frac{GMr(r^2-3 a^2)}{(r^2+a^2)^3}.
\end{equation}
It is interesting to note that in this form $k$ is independent
of boosts along the axis.
This is a consequence of the fact that the Kerr solution is of
Petrov type D and hence the Kerr symmetry axis provides two 
\emph{special tidal directions} for 
ingoing and outgoing trajectories~\cite{mm,bee}.
For $a=0$, the problem reduces to purely radial relative motion in the
exterior Schwarzschild spacetime.

More generally, starting from the generalized Jacobi equation and limiting
our attention to a two-dimensional world as in Section~\ref{sec:hc},
let us consider the equation of relative motion in the form 
\begin{equation}\label{eq:2dw}
\frac{d^2 X}{dT^2}+\kappa (1-2 V^2)X=0,
\end{equation}
where $\kappa(T)$, $\kappa=\fr_{0101}$, is the Gaussian curvature of the surface $(X,T)$ and 
$V=dX/dT$.
\begin{figure}[ht]
\centerline{\epsffile{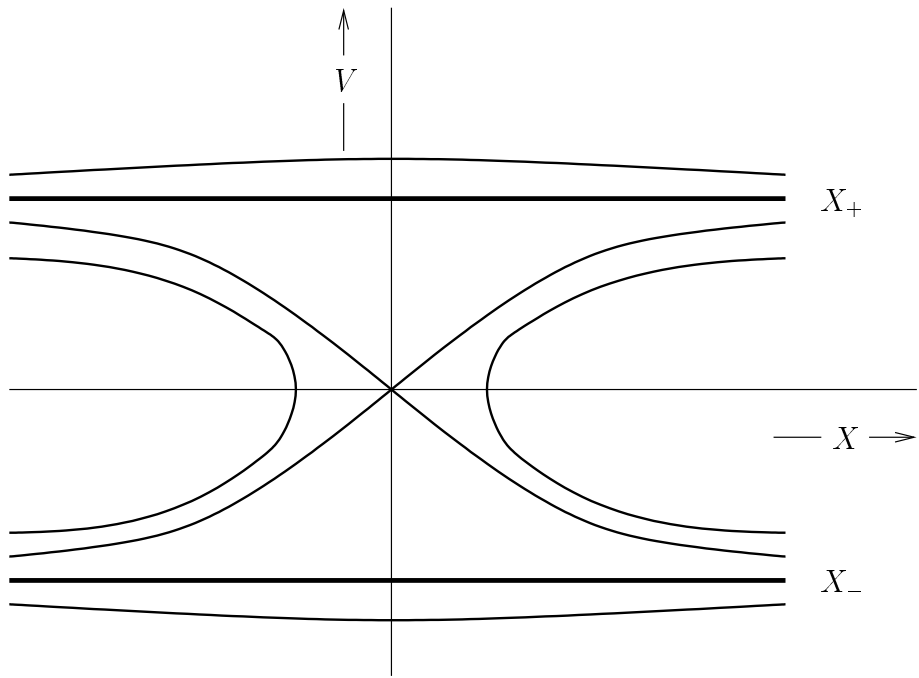}}
\caption{Phase diagram of equation~\eqref{eq:2dw} for negative
constant curvature. The bold horizontal lines correspond to 
$X_{\pm}$.\label{fig2}}
\end{figure}
\begin{figure}[ht]
\centerline{\epsffile{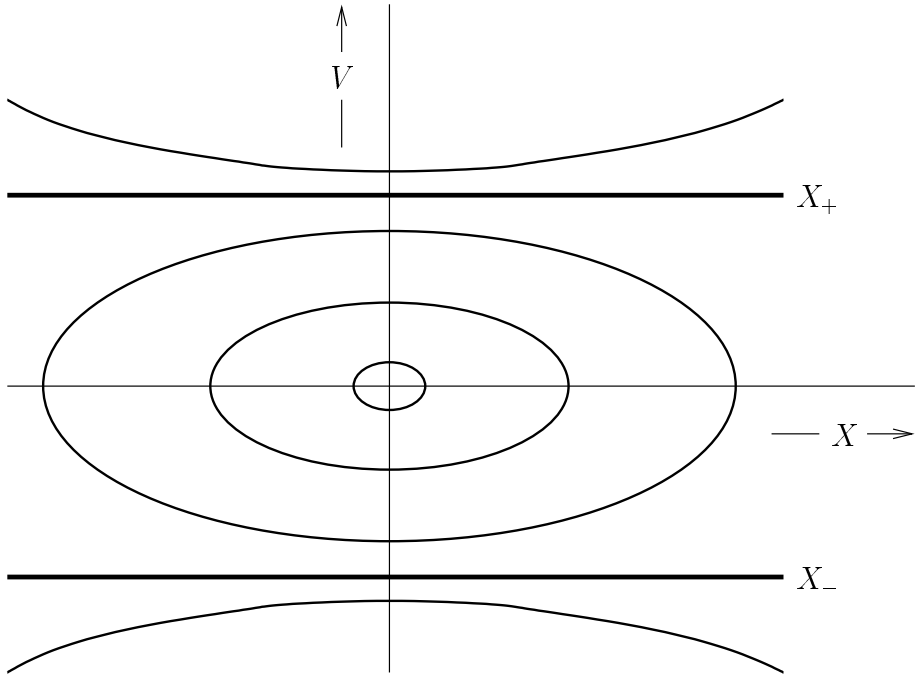}}
\caption{Phase diagram of equation~\eqref{eq:2dw} for  positive
constant curvature. The bold horizontal lines correspond to 
$X_{\pm}$.\label{fig3}}
\end{figure} 

To investigate the main consequences of equation~\eqref{eq:2dw}
beyond the original Jacobi equation for arbitrary curvature,
we assume that $|V|$, $-1<V<1$,
is not negligibly small compared to unity. In particular, let
us note that equation~\eqref{eq:2dw} has constant velocity solutions with 
$V^2=1/2$ regardless of the curvature of the surface, i.e.\
$X_\pm=X_0\pm (1/2)^{1/2}(T-T_0)$, where $T_0$ is a constant initial time and
$X_0=X(T_0)$. It is therefore interesting to investigate the general
behavior of solutions of equation~\eqref{eq:2dw} 
near these constant velocity solutions.
The asymptotic behavior of nearby solutions is determined
by the curvature. Figures~\ref{fig2}
and~\ref{fig3} depict the phase portraits of system~\eqref{eq:2dw} 
for \emph{constant} curvature cases $\kappa<0$ and
$\kappa>0$, respectively.
We note that for constant $\kappa$, equation~\eqref{eq:2dw} is
completely integrable. It would be interesting to investigate the
behavior of the solutions of equation~\eqref{eq:2dw} near $X_\pm$ for
arbitrary $\kappa(T)$. In the rest of this section we limit our
attention to $\kappa(T)<0$, since it follows from equation~\eqref{eq:4.3}
that this is the relevant case for motion away from a Kerr source
($r^2>3 a^2$). 

For the negative curvature case we consider the asymptotic stability
of the invariant sets $\calX_{\pm}:=\{(X,V,T): V=\pm (1/2)^{1/2}\}$ 
in the extended state space for equation~\eqref{eq:2dw}. For example,
we call $\calX_+$  asymptotically stable if  
for every solution $(X(T),V(T),T)$ in the extended state space of 
equation~\eqref{eq:2dw}
that starts near $\calX_+$, 
its distance to $\calX_+$ along the velocity direction 
decreases to zero as $T$ increases to infinity.
It can be shown that asymptotic stability is essentially determined by the 
linearized first-order system along $\calX_+$,  
i.e. the system involving the linear approximation
$W$ of $X-X_+$ and $Q$ of $V- (1/2)^{1/2}$ that is given by
\begin{equation}\label{eq:l2dw}
\frac{dW}{dT}=Q,\qquad \frac{dQ}{dT}=
2\kappa(T) [\sqrt{2}\,X_0+(T-T_0)]Q.
\end{equation}
By integration of system~\eqref{eq:l2dw}, we find that the invariant set
$\calX_+$ of the equation~\eqref{eq:2dw} is positively asymptotically stable  
if for each  real number $C_0$ we have 
\begin{equation}\label{intcond}
\int_{T_0}^\infty \kappa(T)(T+C_0)\,dT=-\infty.
\end{equation}
This result has a useful corollary: 
if $\kappa(T)$ is
asymptotic  to a negative constant multiple
of $T^{-p}$ with $p\le 2$ as $T\to \infty$, then 
$\calX_+$ is asymptotically stable.  

Let us note that if $\gamma<1$ in the differential equation~\eqref{eq:4.1},
then there is a real solution only if the right-hand side of the differential
equation is positive. 
In this case the real solutions with $r>0$ remain 
bounded between
the two positive roots of the right-hand side. 
Under the additional assumption
that the square of the smaller positive root exceeds $3 a^2$, the
curvature obtained from equation~\eqref{eq:4.3} is bounded below zero; 
therefore, the integral condition~\eqref{intcond} is satisfied and
$\calX_+$ is positively asymptotically stable.
For $\gamma=1$, by using an asymptotic analysis of equation~\eqref{eq:4.1}
we find that $r(T)$ is asymptotic to a  constant multiple
of $T^{2/3}$  as $T\to\infty$ and $\kappa(T)$ is therefore asymptotic
to a negative constant multiple of $T^{-2}$. Thus, for $\gamma=1$ the 
invariant set $\calX_+$ is again asymptotically stable; 
the rate of attraction, however,
is slower than any exponential rate. For $\gamma>1$, the 
set $\calX_+$ is no longer asymptotically stable. 
Although nearby solutions in this case
move closer to  $\calX_+$ as $T\to\infty$, 
they are bounded away from this manifold. Let us note that the
results of this analysis also apply to the behavior of solutions near
$\calX_-$, since in this case the linearized system is identical with 
equation~\eqref{eq:l2dw} except for $X_0\mapsto -X_0$. 

The $\kappa<0$ case is particularly interesting in connection with
the problem of ``superluminal'' jets in astrophysics~\cite{fen}. In fact,
the Galactic ``superluminal'' jet sources GRS~1915$+$105 and GRO~J1655$-$40
have speeds that may be  near $(1/2)^{1/2}\approx 0.7$.

\section{Discussion} 
The main dynamical features of the generalized Jacobi equation have 
been presented in this paper. In contrast to the standard Jacobi
equation, the generalized Jacobi equation is nonlinear and therefore
exhibits Hamiltonian chaos under certain circumstances. 
Two specific applications have been considered: 
relative test particle motion in the field of a plane gravitational wave
and in the field of a rotating mass. The results may be relevant
respectively
to future space-based gravitational
wave detectors and high-energy astrophysical phenomena associated with jets.

The bulk speed of Galactic ``superluminal'' jets appears
to be constant and estimated to be around 90\% of the speed of light,
though many uncertainties are associated with
this estimate~\cite{fen}. Jet motion is detected via the temporal
evolution of the radiation intensity contours relative to the
background, which would indicate the motion of blobs (electron clouds)
in the jet relative to the ambient medium. This motion, under the influence
of gravitation alone, is described by the generalized Jacobi equation
as discussed in detail in the previous section. In this case, the
generalized Jacobi equation exhibits an attractor with constant
speed $2^{-1/2}$, approximately equal to 70\% of the speed of light.

Assuming that far from the source gravitational tidal forces are dominant,
it follows from our results that the jet speed should approach $2^{-1/2}$ from 
above and below over a certain characteristic timescale. This timescale
can be calculated from the solution of equation~\eqref{eq:l2dw} for
$Q$. More explicitly, let us consider the jet along the axis of  a Kerr
black hole of mass $M$ and angular momentum $J\le M^2$ at a
distance $r_0\gg 2 GM$; then, the characteristic timescale is $\rho_0$,
where $\rho_0$ is the radius of curvature at $r_0$ given by 
$\rho_0\approx (r_0^3/2 GM)^{1/2}$. 
Simple estimates suggest that further refinements of the observational
techniques~\cite{fen} are required before it may be possible to test this prediction
of the elementary theory of the asymptotic jet speed presented in this paper.

\appendix\setcounter{equation}{0}
\section{Fermi coordinate systems}
Consider a reference observer following a curve $C$ in spacetime.
Let $\tau$ be the proper time along $C$ and $u^\mu=dx^\mu/d\tau$
be its tangent vector. The fiducial observer carries a triad
$\lambda^\mu_{(i)}$ representing three ideal gyroscope directions so
that $\lambda^\mu_{(i)}$ are parallel transported and $\lambda^\mu_{(\alpha)}$
with $\lambda^\mu_{(0)}=u^\mu$ is an orthonormal tetrad frame. 
The Fermi system is a geodesic coordinate system valid in a cylindrical
spacetime region around $C$ such that for any point $P$ with Fermi 
coordinates $X^\alpha=(T, X^i)$ there exists a \emph{unique} 
spacelike geodesic orthogonal to $C$ that connects $P$ with 
a point $P_0$ on $C$.
\begin{figure}[ht]\label{fig:1}
\centerline{\epsffile{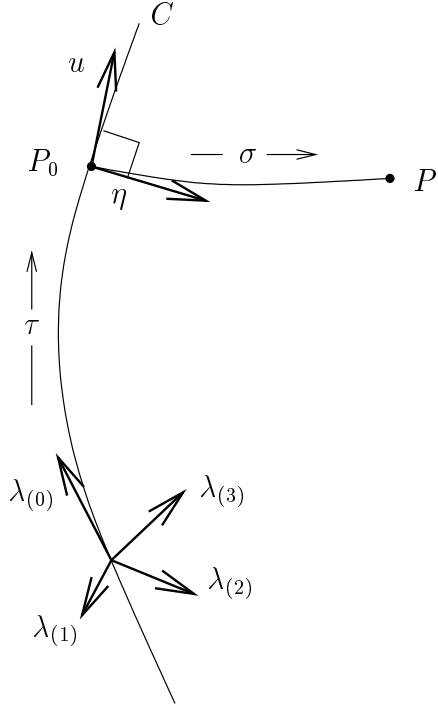}}
\caption{ Schematic diagram depicting the construction of the Fermi normal
coordinate system.}
\end{figure}
Let $\tau$ be the proper time at $P_0$, $\sigma$ be the proper length
of the spacelike geodesic starting from $P_0$ and 
$\eta^\mu=(dx^\mu/d\sigma)_{P_0}$ be the vector tangent to the spacelike
geodesic at $P_0$; then, $\eta_\mu u^\mu=0$ and the Fermi coordinates
are defined by
\begin{equation}\label{eq:a1}
T=\tau,\qquad X^i=\delta^{ij}\sigma\, \eta_\mu\lambda^\mu_{(j)}
\end{equation}
(cf. Figure~3).

To determine the metric tensor $g_{\mu\nu}(x)$ in Fermi coordinates,
i.e.\ $\fg_{\mu\nu}(X)$, we expand $\fg_{\mu\nu}(P)$ in a Taylor series
about $\fg_{\mu\nu}(P_0)=\eta_{\mu\nu}$.
In this expansion, the required derivatives of $\fg_{\mu\nu}$ at
$P_0:(T,\mathbf{0})$ are calculated
on the basis of the following considerations:
The equations of motion of the tetrad frame carried by the observer along
the fiducial geodesic $C$ in Fermi coordinates reduces to 
$\fgam^\mu_{00}(T,\mathbf{0})=0$ and $\fgam^\mu_{0i}(T,\mathbf{0})=0$.
Moreover, the spacelike curve connecting $P_0$ to $P$ satisfies the
geodesic equation, which in Fermi coordinates~\eqref{eq:a1} 
reduces to $\fgam^\mu_{ij}(T,\mathbf{X}) X^i X^j=0$.
Expanding $\fgam^\mu_{ij}$ in a Taylor series about $P_0:(T, \mathbf{0})$,
we find that
\begin{equation}\label{a2}
\fgam^\mu_{ij}(T,\mathbf{0})=0,\quad 
\fgam^\mu_{(ij,k)}(T,\mathbf{0})=0,\quad
\fgam^\mu_{(ij,k\ell)}(T,\mathbf{0})=0,\cdots .
\end{equation}
Using all these relations regarding the connection coefficients at 
$P_0$, the required derivatives of the metric tensor at $P_0$ can
be calculated. The results can be expressed as 
\begin{eqnarray}
\label{eq:a3}
\fg_{00}&=&-1-\fr_{i0j0}X^iX^j-\frac{2}{3}\fcalR_{i0j0k}X^iX^jX^k+\cdots,\\
\label{eq:a4}
\fg_{0i}&=&-\frac{2}{3}\fr_{0jik}X^jX^k
  +\frac{1}{2}\fcalR_{0jki\ell}X^jX^kX^\ell+\cdots,\\ 
\label{eq:a5}
\fg_{ij}&=&\delta_{ij}-\frac{1}{3}\fr_{ikj\ell}X^kX^\ell
  +\frac{1}{3}\fcalR_{i\ell rjk}X^kX^\ell X^r+\cdots,
\end{eqnarray}
where $\calR_{\mu\nu\rho\sigma\omega}$ is given in terms of covariant
derivatives of the Riemann tensor by
\begin{equation}\label{a6}
\calR_{\mu\nu\rho\sigma\omega}=\frac{1}{2}(R_{\mu\nu\rho\sigma;\omega}
+R_{\mu\rho\omega\nu;\sigma}).
\end{equation}
It is simple to show that the tensor $\calR_{\mu\nu\rho\sigma\omega}$
satisfies the following  relations
\begin{equation}\label{a7}
\calR_{\mu\nu\rho\sigma\omega}=-\calR_{\mu\rho\nu\omega\sigma},\quad
\calR_{\mu\nu\rho\sigma\omega}-\calR_{\mu\rho\omega\nu\sigma}
=R_{\rho\sigma\mu(\nu;\omega)}.
\end{equation}
The first covariant derivatives of the Riemann tensor are not completely
specified by the tensor $\calR_{\mu\nu\rho\sigma\omega}$. Moreover,
the next terms in equations~\eqref{eq:a3}--\eqref{eq:a5}
contain, among other terms, the tensor
\begin{equation}\label{a8}
\calT_{\mu\nu\rho\sigma\omega\pi}
=\frac{1}{3}\calR_{\mu\nu\rho\sigma\omega;\pi}
+\frac{2}{3}R^\xi_{\;\rho\omega\nu}R_{\mu\xi\pi\sigma},
\end{equation}
etc. 
It follows from these results that the reduced geodesic 
equation~\eqref{geoiso} in Fermi coordinates becomes the 
\emph{tidal equation}
\begin{equation}\label{a9}
\frac{d^2X^i}{dT^2}+K^i_{\;j}X^j
+\frac{1}{2!}K^i_{\;jk}X^jX^k+\frac{1}{3!}K^i_{\;jk\ell}X^jX^kX^\ell+\cdots=0,
\end{equation}
where the quantities $K_{ij}$, $K_{ijk}$, etc., characterize, respectively,
the tidal acceleration terms of the first order, second order, etc.
We find from equations~\eqref{eq:a3}--\eqref{eq:a5} that
\begin{eqnarray*}
K_{ij}&=&\fr_{0i0j}-2\fr_{0jik}\dot X^k
-2(\fr_{0j0k}\dot X^i-\frac{1}{3}\fr_{i\ell jk}\dot X^\ell)\dot X^k\\
&&{}-\frac{2}{3}\fr_{0\ell kj}\dot X^i\dot X^\ell \dot X^k,
\end{eqnarray*}
as in the generalized Jacobi equation and $K_{ijk}=2K'_{i(jk)}$,
where
\begin{eqnarray*}
K'_{ijk}&=&\fcalR_{i0j0k}+2[\fcalR_{0j\ell ik}
-\frac{1}{3}(\fcalR_{ij\ell k0}+\fcalR_{0ijk\ell})]\dot X^\ell
-\fcalR_{0j0k0}\dot X^i\\
&&{}+(\fcalR_{j\ell i kr}+\frac{5}{6}\fcalR_{kri\ell j})\dot X^\ell\dot X^r
+\frac{2}{3} (2\fcalR_{0j\ell 0k}-\fcalR_{j0k0\ell})\dot X^i\dot X^\ell\\
&&{}+\frac{1}{6}(\fcalR_{k\ell 0jr}
-5 \fcalR_{\ell k0rj})\dot X^i\dot X^\ell\dot X^r.
\end{eqnarray*}

The Jacobi equation in spaces of arbitrary dimensions was first
discussed by Levi-Civita~\cite{lc} and Synge~\cite{s1,s2,s3}.
These references contain a detailed and explicit exposition of the
theorem of Fermi~\cite{f} involving the possibility of choosing a system of
coordinates in a cylindrical region along an arbitrary open curve
in spacetime such that all the connection coefficients vanish and
$g_{\mu\nu}=\eta_{\mu\nu}$ on the curve.

The generalized Jacobi equation was first discussed in~\cite{h,m2,m3}
and developed further in~\cite{ln}. Later, this generalization
was independently rediscovered by Ciufolini~\cite{c1,cd}. Further discussions
of the Jacobi equation can be found, e.g., in~\cite{b,ap, m, k}.
\setcounter{equation}{0}
\section{Hamiltonian equations of motion}
Consider a coordinate system $X^\alpha=(T,X^i)$ in the region
of interest in spacetime such that the metric tensor takes the form
$g_{\mu\nu}=\eta_{\mu\nu}+h_{\mu\nu}$, where $h_{\mu\nu}$ is a small
perturbation on the Minkowski background. In particular, we are
interested in Fermi normal coordinates~\eqref{metrict}.
The isoenergetically reduced Hamiltonian system is given in these
coordinates by 
\begin{equation}\label{eq:b1}
\frac{d X^i}{dT}=\frac{\partial H}{\partial p_i},
\qquad \frac{d p_i}{dT}=-\frac{\partial H}{\partial X^i},
\end{equation}
where
\begin{equation}\label{eq:b2}
H=\frac{g^{0i}}{g^{00}}p_i
\mp\Big(\frac{1+\tilde g^{ij}p_ip_j}{-g^{00}}\Big)^{1/2}.
\end{equation}
To first order in $h_{\mu\nu}$, the Hamiltonian can be expressed as
\begin{equation}\label{eq:b3}
H=\mp\sqrt{1+p^2}(1-\frac{1}{2}h_{00}-\frac{1}{2}\frac{h^{ij}p_ip_j}{1+p^2})
+h^{0i}p_i,
\end{equation}
where $p^2=\delta^{ij}p_ip_j$ and $H$ is a quadratic form in spatial
Fermi coordinates $\mathbf{X}$. Using~\eqref{eq:b3}, the system of 
equations~\eqref{eq:b1} can be written as
\begin{eqnarray}
\label{eq:b4}
\frac{d X^i}{dT}&=&\mp \frac{p_j}{\sqrt{1+p^2}}
[(1-\frac{1}{2}h_{00}+\frac{1}{2}\frac{h^{k\ell}p_kp_\ell}{1+p^2})\delta^{ij}
-h^{ij}]+h^{0i},\\
\label{eq:b5}
\frac{d p_i}{dT}&=&\mp \frac{1}{2}\sqrt{1+p^2}\,
(h_{00,i}+\frac{h^{k\ell}_{\;\;\;,i}\,p_k p_\ell}{1+p^2})
-h^{0j}_{\;\;\;,i}\,p_j.
\end{eqnarray}

Letting $\mp p_i\to \hat p_i$ and defining 
\begin{equation}\label{eq:b6}
P_i:=\frac{\hat p_i}{\sqrt{1+\hat p^2}}
\end{equation}
and $P^i:=\delta^{ij}P_j$, we find that
$(1+\hat p^2)(1-P^2)=1$ and
$\hat p_i=P_i/\sqrt{1-P^2}$.
In terms of $\mathbf{P}$, the Hamiltonian equations of 
motion~\eqref{eq:b4}--\eqref{eq:b5}
take the form
\begin{eqnarray}
\label{eq:b7}
\frac{d X^i}{dT}&=&P^i
(1-\frac{1}{2}h_{00}+\frac{1}{2} h^{k\ell}P_kP_\ell)+h^{0i}\\
\nonumber &&{}
-h^{ij}P_j+\frac{1}{2}h^{k\ell}P_kP_\ell P^i,\\
\label{eq:b8}
\sqrt{1-P^2}\,\frac{d}{dT}\Big(\frac{P_i}{\sqrt{1-P^2}}\Big)
&=& \frac{1}{2}h_{00,i}-h^{0j}_{\;\;\;,i}\,P_j
+\frac{1}{2} h^{k\ell}_{\;\;\;,i}\,P_kP_\ell\,.
\end{eqnarray}

We are interested in the behavior of $X^i=X^i(T)$ to
first order in $\delta=|\mathbf{X}|/\varrho$. To this end,
we note from equation~\eqref{eq:b7} that 
$P^i=V^i+O(h_{\mu\nu})$, where $V^i=dX^i/dT$.
Thus equation~\eqref{eq:b7} can be written to first order in $h_{\mu\nu}$
as 
\begin{equation}\label{eq:b9}
P_i:=(1+\frac{1}{2}h_{00}-\frac{1}{2}h_{k\ell} V^kV^\ell)V_i
+h_{0i}+h_{ij}V^j.
\end{equation}
Similarly, equation~\eqref{eq:b8} can be written as
\begin{equation}\label{eq:b10}
\frac{dP^i}{dT}
+\frac{1}{1-P^2}(\mathbf{P}\cdot\frac{d\mathbf{P}}{dT}) P^i=
\frac{1}{2}h_{00,i}+h_{0j,i}V^j+\frac{1}{2}h_{k\ell,i} V^kV^\ell.
\end{equation}
This relation, after multiplying by $P_i$ and summing over $i$, takes the
form
\begin{equation}\label{eq:b11}
\frac{1}{1-P^2}(\mathbf{P}\cdot\frac{d\mathbf{P}}{dT})=
\frac{1}{2}h_{00,i}V^i+h_{0j,i}V^iV^j+\frac{1}{2}h_{k\ell,i} V^kV^\ell V^i.
\end{equation}
Substituting equations~\eqref{eq:b11} and~\eqref{eq:b9} 
in equation~\eqref{eq:b10} and noting that 
\[
dV^i/dT=O(\delta),\qquad 
h_{\mu\nu,0}=O(\delta^2),
\] 
we finally obtain the 
desired equation of motion
\begin{eqnarray}
\nonumber
\label{eq:b12}
\lefteqn{\frac{dV^i}{dT}+(h_{00,j}V^j+h_{0j,k} V^jV^k) V^i
-\frac{1}{2}h_{00,i}}\\
&&+(h_{0i,j}-h_{0j,i})V^j
+(h_{ij,k}-\frac{1}{2} h_{jk,i})V^jV^k=0.
\end{eqnarray}

Taking due account of the approximations under consideration here, we 
note that
\[
-\Gamma^0_{\alpha\beta}\frac{dX^\alpha}{dT}\frac{dX^\beta}{dT}
=h_{00,j} V^j+h_{0j,k} V^jV^k
\]
and
\begin{eqnarray*}
\Gamma^i_{\alpha\beta}\frac{dX^\alpha}{dT}\frac{dX^\beta}{dT}
&=&\Gamma^i_{00} +2 \Gamma^i_{0j} V^j +\Gamma^i_{jk} V^jV^k\\
&=& -\frac{1}{2}h_{00,i}+(h_{0i,j}-h_{0j,i})V^j
+(h_{ij,k}-\frac{1}{2}h_{jk,i})V^jV^k
\end{eqnarray*}
in agreement with equation~\eqref{geoiso}. 
Moreover, to first order in $\delta$
\begin{eqnarray*}
h_{00,i}&=& -2 R_{0i0j} X^j,\\
h_{0i,j}&=& -\frac{2}{3}( R_{0jik}+ R_{0kij})X^k,\\
h_{\ell m,i}&=& -\frac{1}{3}( R_{\ell imk}+ R_{mi\ell k})X^k,
\end{eqnarray*}
in the Fermi system. 
Substituting these relations in equation~\eqref{eq:b12} and
using the symmetries of the Riemann tensor, such as 
$R_{0[ijk]}=0$, we obtain equation~\eqref{gje}. 
This is the generalized Jacobi equation in Fermi coordinates; for 
arbitrary coordinates see Appendix~C. 
\setcounter{equation}{0}
\section{Generalized Jacobi equation in arbitrary coordinates}
Consider a reference timelike geodesic $C$ as in Appendix~A
representing the worldline of an observer with proper time
$\tau$. Let $\xi^\mu=\sigma\eta^\mu$; then, to first order in 
$\sigma/\varrho$,
the generalized Jacobi equation~\cite{h, m2} takes the form
\begin{eqnarray*}
\nonumber
\lefteqn{
\frac{D^2\xi^\mu}{D\tau^2}
 +R^\mu_{\;\;\rho\nu\sigma} u^\rho\xi^\nu u^\sigma
+(u^\mu+\frac{D\xi^\mu}{D\tau})(2 R_{\zeta\rho\nu\sigma}
u^\zeta\frac{D\xi^\rho}{D\tau}\xi^\nu u^\sigma}\hspace{2in}\\
\nonumber
&&\hspace{-2in}+\frac{2}{3}R_{\zeta\rho\nu\sigma}u^\zeta
\frac{D\xi^\rho}{D\tau}\xi^\nu
\frac{D\xi^\sigma}{D\tau})
+2 R^\mu_{\;\;\rho\nu\sigma}\frac{D\xi^\rho}{D\tau}\xi^\nu u^\sigma
+\frac{2}{3}R^\mu_{\;\;\rho\nu\sigma}\frac{D\xi^\rho}{D\tau}\xi^\nu
  \frac{D\xi^\sigma}{D\tau}=0.\\
\end{eqnarray*}
Here 
\[
\frac{D\xi^\mu}{D\tau}=
\xi^\mu_{\;\; ;\nu}\frac{dx^\nu}{d\tau}=\xi^\mu_{\;\; ;\nu} u^\nu
\]
is the covariant derivative of $\xi^\mu$ along the reference worldline.
If we now use Fermi coordinates and set 
\[\xi^\mu=X^i\lambda^\mu_{(i)},\quad 
\frac{D\xi^\mu}{D\tau}=\frac{dX^i}{dT}\lambda^\mu_{(i)},\quad
\frac{D^2\xi^\mu}{D\tau^2}=\frac{d^2X^i}{dT^2}\lambda^\mu_{(i)},\ldots,
\]
we recover equation~\eqref{gje}. Here we have used the
relation $D\lambda^\mu_{(\alpha)}/D\tau=0$; i.e.\ 
the tetrad is parallel transported along $C$.
\section{Worldlines of observers at rest}
Consider a spacetime metric of the form
\[
ds^2=-dt^2 +g_{ij}(t, x^k) dx^i dx^j.
\]
If a particle is a rest at a point in the corresponding space, then its 
worldline follows a geodesic. Indeed, 
the four-velocity of the particle $u^\mu=dx^\mu/d\tau$ is given
by $u^0=1$ and $u^i=0$. We must show that
\[ \frac{du^\mu}{d\tau}+\Gamma^\mu_{\alpha\beta}u^\alpha u^\beta=0.\]
For $\mu=0$, it suffices to prove that $\Gamma^0_{00}=0$. 
Because of the block form of the metric, we have $g^{00}=-1$ and $g^{0i}=0$; 
therefore,
\[\Gamma^0_{00}=-\frac{1}{2}g_{00,0}=0.\] 
For $\mu=i$, it suffices to show that $\Gamma^i_{00}=0$.
In this case we have $g_{0j}=0$ and $g_{00}=-1$. Hence,
\[\Gamma^i_{00}=\frac{1}{2}g^{ij}(2 g_{j0,0}-g_{00,j})=0,\]
as required.


\begin{thebibliography}{xxxx}
\bibitem{s3} J. L. Synge, Relativity: The General Theory (North-Holland,
Amsterdam, 1960).
\bibitem{b1} B. Mashhoon, Proc. Second Marcel Grossmann Meeting
on General Relativity, edited by R. Ruffini 
(North-Holland, New York, 1982) pp. 647--672; N.B. section IV.
\bibitem{bee} J.\ K.~Beem, P.\ E.~Ehrlich and K.\ L.~Easley,
Global Lorentzian Geometry, 2nd ed. (Dekker, New York, 1996) ch.~2.
\bibitem{k} R. Kerner, J. W. van Holten and R. Colistete, Jr., Class. Quantum
Grav. 18 (2001) 4725.
\bibitem{am} R. Abraham and J. Marsden, Foundations of Mechanics,
2nd Edition (Perseus Books, Reading, 1985) pp. 223--225.
\bibitem{bpr} H. Bondi, F. A. E. Pirani, and I. Robinson,
Proc. Roy. Soc. A 251 (1959) 519.
\bibitem{lisa} The LISA mission is described in 
http://lisa.jpl.nasa.gov/ and 
http://www.estec.esa.nl/spdwww/future/html/lisa.htm.
\bibitem{mm} B. Mashhoon and J.\ C.~McClune, Mon. Not. R. Astron. Soc.
262 (1993) 881.
\bibitem{fen} R. Fender, astro-ph/0109502.
\bibitem{lc} T. Levi-Civita, Math. Ann. 97 (1926) 291.
\bibitem{s1} J. L. Synge, Phil. Trans. Roy. Soc. London A 226 (1926) 31.
\bibitem{s2} J. L. Synge, Duke Math. J. 1 (1935) 527.
\bibitem{f} E. Fermi, Atti Accad. Naz. Lincei Rend. Cl. Sci. Fis. Mat.
Nat. 31 (1992) 21, 51. 
\bibitem{h} D. E. Hodgkinson, Gen. Rel. Grav. 3 (1972) 351.
\bibitem{m2}  B. Mashhoon, Astrophys. J. 197 (1975) 705.
\bibitem{m3}  B. Mashhoon, Astrophys. J. 216 (1977) 591.
\bibitem{ln} W.-Q. Li and W.-T. Ni, J. Math. Phys. 20 (1979) 1473, 1925.
\bibitem{c1} I. Ciufolini, Phys. Rev. D 34 (1986) 1014.
\bibitem{cd} I. Ciufolini and M. Demia\'nski, Phys. Rev. D 34 (1986) 1018.
\bibitem{b} S. L. Ba\.za\'nski, 
  Ann. Inst. Henri Poincar\'e 27 (1977) 115, 145.
\bibitem{ap} A.\ N.~Alexandrov and K.\ A.~Piragas, 
              Theoret.~Math.~Phys. 38 (1979) 48.
\bibitem{m} S. Manoff, Int.\ J.\ Mod.\ Phys.\ A 16 (2001) 1109.
\end{thebibliography}
\end{document}